\def \cm{~\rm{cm}}
\def \s{~\rm{s}}
\def \km{~\rm{km}}

\def \K{~\rm{K}}
\def \g{~\rm{g}}

\def \AU{~\rm{AU}}
\def \erg{~\rm{erg}}

\def \yr{~\rm{yr}}

\documentclass[12pt,preprint]{aastex}
\usepackage{natbib}

\shorttitle{CFW Model for Eta Car}
\shortauthors{Soker}
\slugcomment{Draft version of \today}

\begin{document}

\title{Why a Single-Star Model Cannot Explain the Bipolar Nebula of
Eta Carinae}

\author{Noam Soker}
\altaffiltext{1}
{Department of Physics, Technion$-$Israel Institute of Technology,
Haifa 32000 Israel, and Department of Physics, Oranim, Israel;
soker@physics.technion.ac.il.}

\begin{abstract}

I examine the angular momentum evolution during the 1837-1856
Great Eruption of the massive star $\eta$ Carinae. I find
that the new estimate of the mass blown during that eruption
implies that the envelope of $\eta$ Car substantially spun-down
during the 20 years eruption. Single-star models, most of
which require the envelope to rotate close to the break-up
velocity, cannot account for the bipolar nebula$-$the
Homunculus$-$formed from matter expelled in that eruption.
The kinetic energy and momentum of the Homunculus further constrains
single-star models. I discuss how $\eta$ Car can fit into a
unified model for the formation of bipolar lobes where two oppositely
ejected jets inflate two lobes (or bubbles). These jets are blown by
an accretion disk, which requires stellar companions in the case
of bipolar nebulae around stellar objects.

\end{abstract}


\section{Introduction}
\label{sec:intro}
 The bipolar structure of the Eta Carinae ($\eta$
Car) nebula$-$the Homunculus$-$is not unique.
 Its basic structure, that of bubble pair, where each bubble is
bounded by a thin dense shell, filled with a low density interior,
and having a narrow waist between them, is seen in many diverse
objects.
 Examples include the Perseus cluster of galaxies
(Fabian et al.\ 2002), the symbiotic nebula He 2-104 (Corradi \&
Schwarz 1995; Corradi et al. 2001; here the dense shell is not
closed), and the planetary nebula (PN) NGC 3587 (PN G148.4+57.0;
e.g., Guerrero et al.\ 2003).
 The similarity between bubble pairs in clusters of galaxies and
in PNs is discussed in Soker (2003a,b; 2004a), while in Soker (2004b)
I discuss the similarity of these systems with symbiotic nebulae and
the bipolar nebula of $\eta$ Car.
 In clusters of galaxies such bubbles are known to be formed by
oppositely ejected jets, which are detected by radio emission
(e.g., Hydra A; McNamara et a.\ 2000).
 In a series of papers I argue that this suggests that bubble pairs
in PNs (Soker 2003a,b; 2004a) and symbiotic nebulae (Soker 2004b) are
also formed by such jets (note that not all PNs are shaped by
jets).
 In particular, there are more and more observations showing and hinting
at the presence of jets in symbiotic systems (e.g., Kellogg,
Pedelty, \& Lyon 2001; Brocksopp et al.\ 2004), in PNs, and PN
progenitors (e.g., Imai et al.\ 2002; Hirano et al.\ 2004; Sahai
et al.\ 2003).

In previous papers I attribute the bipolar structure of $\eta$ Car
(e.g., Ishibashi et al.\ 2003) and the presently blown fast polar
wind found by Smith et al.\ (2003a) to interaction with the binary
companion (Soker 2001, 2003c).
 In particular, in Soker (2001) I proposed that the bipolar
nebula was shaped by jets blown by the companion, via an accretion
disk, during the 20 years Great Eruption a century and a half ago.
In that paper I listed some arguments in favor of such a model.
However, the new finding of Smith et al.\ (2003b) of a more
massive Homunculus, and several new papers using a single star
model for the shaping of the wind and circumstellar matter of
$\eta$ Car (e.g., Dwarkadas \& Owocki 2002; Smith et al.\ 2003a;
Gonzalez et al.\ 2004; van Boekel et al.\ 2003) motivate me to
reconsider the single star model.
{{{ Other papers in recent years study the role of rotation and/or binary
companion in $\eta$ Car. Stothers (1999) find that the rotation does
not affect much the instability of luminous blue variables (LBV),
which is in support of the view I take in the present paper.
Davidson (1999; see also Davidson 2000), on the other hand, argues that
a binary model for $\eta$ Car is not ``a Panacea''.
I accept Davidson's view that some crucial aspects of the $\eta$ Car
eruptions are due solely to the behavior of the massive star, and
not to the companion.
My main interest is in showing that the bipolar structure of the nebula
is due to a binary companion. }}}
In section 2 I study the momentum and energy budget of $\eta$ Car.
In section 3 I consider the required stellar angular momentum in
single-star models.
{{{ I will not follow the evolution of rotating massive stars with
any specific model (for these see, e.g., Heger \& Langer 2000;
Maeder, \& Meynet 2003;  Meynet \& Maeder 2003).
I will use analytical approach. }}}
In section 4 I summarize my finding that single-star models
encounter severe problems.

\section{Wind's energy and momentum}
\label{sec:energy}
 Some of the arguments given here are not new.
However, I put them in a broader context in light of
the new estimated mass that was blown during the 20 years Great
Eruption $M_{\rm GE} \simeq 12 M_\odot$ (Smith et al.\ 2003b).
 For a wind driven by radiation pressure the effective number of times
a photon is scattered by the wind material, i.e., in the positive
radial direction, is given by
\begin{equation} \eta_s \simeq \frac{\dot M_w v_w}{L/c} = 700
\left(\frac{L_\ast}{10^{7.3} L_\odot} \right)^{-1}
\left(\frac{\dot M}{0.5 M_\odot \yr^{-1}} \right)
\left(\frac{v_w}{600 \km \s^{-1}} \right),
\end{equation}
where $\dot M_w$ is the mass loss rate into the wind, $v_w$ is the
terminal wind speed, $L_\ast$ is the stellar luminosity, and $c$
the light speed. The different variables are scaled with the mass
loss rate (Smith et al.\ 2003b) and luminosity (Davidson \&
Humphreys 1997) during the Great Eruption.
This is a very large number, which is not encountered in other stars.
 For example, in  most PNs and highly evolved red giant stars
$n_s <1$ (Knapp 1986).
 The PN with the highest value of $\eta_s \sim 600$ in the list given
by Kanpp (1986) is NGC 2346. The central star of this PN contains a
close binary companion (Bond 2000), which went through a common
envelope evolution and released large amount or orbital energy.
Therefore, the  binary companion is behind the large value of $\eta_s$
in the PN NGC 2346.
 Smith et al.\ (2003b) notice that such a high value of $\eta_s$
occurs in explosions. However, the mass loss rate during the
Great Eruptions lasted for $\sim 20 $ years, a time longer by more than
an order of magnitude than the dynamical time of $\eta$ Car, even
if it swelled to $\sim 10 \AU$.
 Shaviv (2001) compares the super-Eddington wind of the Great
Eruption to that of novae and obtains satisfactory fit to the Great
Eruption outflow, assuming spherical wind.
Although he assumes a steady state mass loss process, the
photospheric radius of the nova ejected mass expands by several
orders of magnitude during the explosion, and the mass involved
occupies a thin layer on the white dwarf surface prior to the
explosion.
The $\sim 10 M_\odot$ ejected mass in the Great Eruption, on the
other hand, originated in a thick envelope layer of $\eta$ Car.
In addition, the formation of the bipolar structure in an explosive
event requires different models than the one I criticize in the
present paper.
The results of Shaviv (2001), though, show that a single star can
lose mass at a high rate. Such a high mass loss rate is needed
in the presently proposed binary model, but here the
wind speed, hence the momentum and kinetic energy supplied by
the primary, can be much lower.

 The new estimated energy of the Homunculus is
$E_{\rm Hom} \sim 10^{49.6}-10^{50} \erg$ (Smith et al.\ 2003b).
The ratio of this energy to the energy radiated during the
20 years Great Eruption is
\begin{equation}
\eta_E \equiv \frac{E_{\rm Hom}}{L_\ast \times 20\yr}
=1     \left(\frac{L_\ast}{10^{7.3} L_\odot} \right)^{-1}
\left(\frac{E_{\rm Hom}}{5 \times 10^{49} \erg} \right)
\end{equation}
This implies that half the energy liberated during the Great
Eruption was radiated, and half was converted to the kinetic
energy of the wind.
The present mass loss rate of $\lesssim 2
\times 10^{-3} M_\odot \yr^{-1}$ (van Boekel et al.\ 2003) over
the $\sim 150$ years span since the Great Eruption contributes
negligible amount to this energy.

The required transfer efficiency of momentum and energy from radiation
to the wind in single star models of $\eta$ Car is much higher
than in any other object blowing similar winds.
These values are more typical for binary systems, as I now discuss.
In principle there are two processes by which a binary companion
can account for such a high kinetic energy of the wind.
First, the binary system can release orbital energy
$\Delta E_{\rm orb} \simeq 0.5 G M_1 M_2 /a_f$ where
$M_1$ and $M_2$ are the two masses inward to the final orbital
separation of $a_f$, and assuming that the final orbital separation
is much smaller than the initial orbital separation.
If a fraction $\chi$ of the released orbital energy is deposited
into the expelled mass, then the final orbital separation for a
companion  to explain the Homunculus energy is
\begin{equation}
a_f \lesssim 70 \left( \frac{E_{\rm Hom}}{5 \times 10^{49} \erg}
\right)^{-1}
\left( \frac{\chi}{0.5} \right)
\left( \frac{M_1}{120 \time 30 M_\odot} \right)
\left( \frac{M_2}{30 \time 30 M_\odot} \right) R_\odot.
\end{equation}
 The presently observed orbital period is $5.5 \yr$, which for
primary and companion masses of $M_1 = 120 M_\odot$ and
$M_2=30 M_\odot$, respectively, implies an average orbital separation of
$a \sim 16.5 \AU$ (for observational support for
the presence of a binary and its properties see, e.g., Damineli
1996; Ishibashi et al.\ 1999; Damineli et al.\ 2000; Corcoran et
al.\ 2001a,b; Pittard \& Corcoran 2002; Duncan \& White 2003
Fernandez Lajus et al.\ 2003).
 Therefore, the orbital energy of the presently observed binary system
cannot account for the kinetic energy of the Homunculus.
However, $\eta$ Car could have swallowed a closer, lower mass
third star during its Great Eruption.
For example, a third star with $M_3 = 5 M_\odot$ could have
spiral down to the core of $\eta$ Car, and released more than the
required energy. The Lesser Eruption of 1890 makes this scenario
unlikely, although it can't be ruled out based on pure physical
arguments.

In the second process, the companion is outside the primary
envelope, it accretes mass via an accretion disk and blows two
jets (or a collimated fast wind: CFW). In this scenario, which was
proposed for the formation of the Homunculus in Soker (2001; where
more supporting arguments are given), the primary in $\eta$ Car
expelled its mass at low speeds.
 In principle, the companion may accretes half of the mass expelled by
the primary and blow a fraction of $\sim 0.2$ at a speed equals to
its escape velocity $v_j \sim 2000 \km \s^{-1}$.
For an accreted mass of $\sim 10 M_\odot$ the energy carried in
the jets is then $\sim 8 \times 10^{49} \erg$, enough to account
for the energy of the Homunculus.
In addition, in such a scenario a substantial fraction of the
energy radiated by the system during the great eruption came from the
accretion energy onto the companion.
Recently, Smith \& Morse (2004) reported the discovery
of extremely fast material, $v> 3200 \km \s^{-1}$, in $\eta$ Car.
They attribute this material to mass loss from the primary star.
In the binary model, on the other hand, this extremely fast material
was first accreted onto an accretion disk around the companion,
and then was ejected at this high speed.

\section{The angular momentum evolution of $\eta$ Car}
\label{sec:angular}

\subsection{Moment of Inertia}

 Following Soker \& Harpaz (1999)
\footnote{Note that the density scale in Figs.\ 1-5 of SH99 is too
low by a factor of 10; the correct scale is displayed in their
Fig. 6.} I examine the ratio of the density
at the photosphere, $\rho_p$, to the average envelope density $\rho_a$.
As shown below, this ratio indicates the moment of inertia, a relevant
quantity for the slowing down process.
The photospheric density is given by (Kippenhahn \&
Weigert 1990)
\begin{equation}
\rho_p = \frac{2}{3} \frac {\mu m_H}{k_B}
 \frac {G M_1}{R^2 \kappa T},
\end{equation}
where $\mu m_H$ is the mean mass per particle, $k_B$ is the
Boltzmann constant, and $\kappa$ is the opacity. Substituting
typical values for $\eta$ Car in the last equation gives
\begin{equation}
\rho_p= 1.5 \times 10^{-13} \left( \frac{\kappa}{1 \cm^2 \g^{-1}}  
\right)^{-1} \left( \frac{T}{10^4 \K} \right)^{3} \left(
\frac{L_1}{10^7 L_\odot} \right)^{-1} \left( \frac{M_1}{120
M_\odot} \right)  \g \cm^{-3},
\end{equation}
where a black body luminosity was assuemd to eliminate the stellar
radius.
 Using the (solar composition) opacity as given by
{{{ Alexander \& Ferguson 1994; see also }}}
Rogers \& Iglesias 1992), I find the following
{{{ approximate, but }}} adequate, fitting in the
relevant photospheric densities
\begin{eqnarray}
 \kappa \simeq \left\{
\begin{array}{cl}
10 (T/10^4 \K)^{14} \cm^2 \g^{-1} & {\rm for} \ 5,000 K \lesssim T \lesssim 8,500 K \\
1 \cm^2 \g^{-1} & {\rm for} \ \quad 8500K \lesssim T .
\end{array}
 \right .
\end{eqnarray}

I take the envelope mass $M_{\rm env}$
to contain about half of the $\eta$ Car stellar mass.
{{{ For LBV as $\eta$ Car, the envelope mass is expected to
be lower, hence increasing the slowing down process and
strengthening the effect studied here. }}}
 The average density in the envelope is given by
\begin{equation}
\rho_a= \frac{M_{\rm con}}{4 \pi R^3/3} \simeq  7 \times 10^{-8} 
\left( \frac{T}{10^4 \K} \right)^{6} \left( \frac{L}{10^7 L_\odot}
\right)^{-3/2} \left( \frac{M_{\rm env}}{60 M_\odot} \right)  \g
\cm^{-3}.
\end{equation}
As was shown for AGB stars (Soker \& Harpaz 1999), and can be
checked for more massive stars, when $\rho_p$ is not much smaller
than $\rho_a$, the envelope density profile is shallow. In low
mass envelopes, the density profile is almost flat in the outer
region (Soker \& Harpaz 1999). For massive stars this can be seen
in the model of an initial $120 M_\odot$ star which was reduced to
$66.6 M_\odot$ as calculated by Stothers \& Chin (1993). In that
model (their fig. 2), the effective temperature is $T_p = 8000
\K$, and luminosity $2 \times 10^6 L_\odot$. The radius is $\sim
700 R_\odot$.
 The envelope mass is very low, because of the mass lost by the
star. On average, the density profile from $\sim 0.05 R_\ast$ to
the surface is $\rho \propto r^{-1}$. The parameters in the
evolved $120 M_\odot$ model of Maeder (1981) without mass loss are
$L \simeq 10^6 L_\odot$, $T_p \simeq 4000 \K$,
$R \simeq 2000 R_\odot$, and $M_{\rm env} = 30 M_\odot$.
 From the equations above, and using the correct opacity for low
 densities of $\kappa \sim 3 \times 10^{-4}$ I find
 $\rho_p = 10^{-10} \g \cm^{-3}$  and
 $\rho_a = 5 \times 10^{-9} \g \cm^{-3}$. The ratio of 50 is moderate, and
 the density profile in the outer envelope is $\rho_e \propto r^{-3}$.

The structure of the envelope considered above determines the
moment of inertia of the star (that of the core is negligible for
giants)
\begin{equation}
I = \alpha M_{\rm env} R^2.
\end{equation}
 For an envelope density profile of
 $\rho_e \propto r^{-2}$ one finds  $\alpha = 2/9$, while for
 $\rho_e \propto r^{-3}$ the value is
 $\alpha = [3 \ln (R/r_{\rm in})]^{-1}$,
 where $r_{\rm in}$ is the inner radius of the envelope.
In the model described above of Maeder (1981),
 $R/r_{\rm in}\simeq 10$, and I find $\alpha \simeq 0.15$.
 For steeper density profiles the value of $\alpha$ is lower.
In the sun, for example, equations (5)-(7) give a very large ratio
of $\log (\rho_a/\rho_p) \simeq 5$. The density profile in most
of the solar interior (beside the very outer envelope) can be
fitted by
\begin{equation}
\rho (r) = \rho_0 \exp (-K r),
\end{equation}
with $K R_\odot$= 10.54 (Bahcall, Pinsonneault, \& Basu 2001). For
such a profile, and with $r_{\rm in} \ll R$, the moment of inertia
coefficient is $\alpha \simeq 8/(KR)^2$.
For the sun this gives $\alpha=0.07$.

This subsection shows that the ratio $\rho_a/\rho_p$ can be used
as a crude indicator for the moment of inertia. The exact value of
$\alpha$ can't be predicted from this ratio, but for the present
goal it is enough to use the crude relation
\begin{eqnarray}
 \alpha \sim \left\{
\begin{array}{cl}
 0.2, &  {\rm for} \ \log(\rho_a/\rho_p) \lesssim 2 \\
 0.1-0.2 & {\rm for} \ 2 \lesssim \log(\rho_a/\rho_p) \lesssim 4 \\
 \lesssim 0.1 & {\rm for} \ 4 \lesssim \log(\rho_a/\rho_p) .
\end{array}
 \right .
\end{eqnarray}

\subsection{Angular Momentum Loss}

Consider an envelope of a giant star rotating as a solid body.
 The angular momentum loss rate from the envelope to the wind is
\begin{eqnarray}
\dot J_{\rm wind} = \beta \omega R^2 \dot M,
\end{eqnarray}
where $\omega$, is the stellar angular velocity, $J$ is the
stellar angular momentum, and $\beta$ depends on the mass loss
geometry: for a constant mass loss rate per unit area on the surface $\beta =2/3$,
while for an equatorial mass loss $\beta=1$.
Smith et al.\ (2003b) argue for an enhanced polar mass loss rate during the
Great Eruption, for which $\beta < 2/3$.
However, still some extra mass resides in the equatorial plane even according
to Smith et al.\ (2003b).
 Therefore, I will take $\beta \simeq 0.5-0.7$ for the Great Eruption
mass loss geometry.
The change of the envelope's angular momentum with mass loss is given by
(e.g., Soker \& Harpaz 2000)
\begin{eqnarray}
\frac {d \ln J_{\rm env}}{d \ln M_{\rm env}} = \frac {d \ln
\omega}{d \ln M_{\rm env}} + \frac {d \ln I}{d \ln M_{\rm env}} =
\frac {\beta}{\alpha (M_{\rm env})} \equiv \delta .
\end{eqnarray}
 If the structure of the atmosphere does not change much while mass loss
occurs then ${d \ln I}/{d \ln M_{\rm env}} = 1$ and $\delta$ is constant.
The solution of the last equation becomes
\begin{eqnarray}
\frac {\omega}{\omega_0} =\left( \frac {M_{\rm env}}{M_{\rm env0}}
\right)^{\delta-1}.
\end{eqnarray}

\subsection{Angular Velocity Evolution During the Great Eruption}

Most single-star models I am aware of for the formation of the
Homunculus require $\eta$ Car to rotate at
 $\Omega \equiv (\omega/\omega_{\rm Kep}) \gtrsim 0.7$,
where $\omega$ is the angular velocity of the stellar envelope
and $\omega_{\rm Kep}$ is the Keplerian
velocity on the equator (i.e., the break-up angular velocity).
Dwarkadas \& Owocki (2002) and Smith et al.\ (2003a), for example, take
$\Omega = 0.9$, while Maeder \& Desjacques (2001) take $\Omega =0.8-0.9$.
The model presented by Langer, Garc\'{\i}a-Segura, \& Mac Low (1999)
is different in that they consider the ratio of luminosity to the
Eddington limit.
{{{ Similar considerations are qualitatively presented by
Zethson et al.\ (1999) to account for the slow ejecta in the
equatorial plane of $\eta$ Car.
They emphasize that even slowly rotating luminous stars can
have highly non-spherical mass loss geometry.
They also point out that their model cannot explain why
the slow ejecta are in dense compact condensations.
In the binary model, these condensations are assumed
to be formed by the influence of the companion. }}}
The model of Langer et al.\ (1999) which gives two lobes similar
to those in $\eta$ Car have $\Omega \lesssim 0.2$.
However, most of the mass in their model
is being lost at low velocities,$\lesssim 400 \km \s^{-1}$, and it
resides in the equatorial plane rather than in the Homunculus.
Therefore, the total kinetic energy in their model is much below
that of the Homunculus.
{{{ Aerts, Lamers, \& Molenberghs (2004)
study the influence of rotation on the mass loss
from $\eta$ Car, and also find the wind velocity to be too low. }}}
Therefore, in what follows I will refer
only to the models require $\Omega \gtrsim 0.7$.

 To facilitated an analytical treatment that will demonstrate the
problems of a rotating single star, I consider two extreme cases:
an eruption accompanied by expansion of the envelope to $R \sim 10 \AU$
(e.g., Davidson \& Humphreys 1997), and an eruption leaving the
$\eta$ Car primary star a hot star.
The angular momentum of $\eta$ Car at the beginning of the Great
Eruption is determined by both single star evolution and binary
interaction.
To posses fast rotation, though, the star must have
interacted with a stellar companion, because single stars slow down with
mass loss (Maeder \& Meynet 2000),
{{{ e.g., large mass loss events before the Great Eruption,
such as the one proposed by Bohigas et al.\ (2000),
would have substantially slowed down $\eta$ Car. }}}
As a star like $\eta$ Car expands to $R \sim 10 \AU$, its moment
of inertia increases both because of the increase in size and
the increase in moment of the inertia coefficient $\alpha$ (eq. 8).
Presently, $\eta$ Car has a radius of $\sim 100 R_\odot$ and
an effective temperature of $\sim 20,000 - 30,000\K$ (Hillier et al.
2001; Smith et al\ 2003a).
 For such a model $\alpha \lesssim 0.1$ by equations (5), (7) and (10).
 For a Great Eruption luminosity of $10^{7.3} L_\odot$ and a
radius of $\sim 10 \AU$, the effective temperature becomes
$\sim 8,300 \K$, and I find $\alpha \sim 0.1$.
Such an expansion implies, because of conservation of angular momentum,
that the envelope must rotate very slowly, even if before expnation the star
was almost at break-up angular velocity.
I take now the following parameters:
total mass lost during the Great Eruption equal to
$\Delta M_{\rm GE} = 12 M_\odot$;
an envelope mass of about half the stellar mass, or
$M_{\rm env0} \sim 70 M_\odot$ at the beginning of the great eruption
{{{ (as stated above, this is an upper limit on the envelope mass;
the envelope mass is likely to be lower, increasing the effect studied
here); }}}
mass loss geometry with $\beta=0.5$
(see discussion following eq. 11); and $\alpha=0.1-0.15 $.
For these parameters, $\delta-1 = (\beta/\alpha)-1 = 4$.
By equation (13), the envelope angular velocity at the end of the
Great Eruption is $\Omega_f=\Omega_0 (58/70)^4= 0.5 \Omega_0$,
where $\Omega_0$ is the (non-dimensional) angular velocity before mass
loss starts but after expansion.
At the middle of the Great Eruption,
$\Omega_m=\Omega_0 (64/70)^4= 0.7 \Omega_0$.
For $\alpha=0.15$, as in the extended $60 M_\odot$ model
of Maeder (1981), these values are
$\Omega_f=  0.65 \Omega_0$ and $\Omega_m= 0.8 \Omega_0$.
Since in this scenario the angular velocity at the beginning
of the Great Eruption should be very low as well $\Omega_0 \ll 1$,
I conclude that in the case when $\eta$ Car expanded during the
Great Eruption the angular velocity is very low, and a single star model
cannot account for the bipolar structure.

Consider now the other extreme, where $\eta$ Car remained a hot star
during the Great Eruption.
Equations (5) (7) and (10) implies $\alpha \lesssim 0.1$.
Again, I take $\beta=0.5$, i.e., a moderate polar enhanced mass loss rate.
The value of $\beta$ cannot be too low. This is because a low value of $\beta$
implies highly enhanced mass loss rate along polar directions. This, however,
will make the problem of the radiation momentum transfer to the wind
(eq. 1), even more severe, as only radiation escaping along the
polar directions accelerate the mass there.
Therefore, a value of $\delta=4$, as in the extended envelope case, may
be an underestimate. In any case, for this value of $\delta$, and for a maximum,
possible value of $\Omega_0=1$, the angular velocity and middle and end
of the Great Eruptions are $\Omega_m=0.7$ and $\Omega_f=0.5$, respectively.

The main conclusion of this section is that as a single star, $\eta$ Car could
not have maintained a fast rotation during the 20 years Great Eruption.

\section{Summary}
\label{sec:summary}

Most single-star models for the formation of the bipolar nebula$-$the
Homunculus$-$of $\eta$ Car are based on fast rotation of the progenitor
during the 20 years Great Eruption, 1843-1856.
These models do not consider the origin of the fast rotation, or the way
the envelope maintains its fast rotation during the Great Eruption.
A single-star spin-up mechanism as proposed by Heger \& Langer (1998)
requires the outer convective part of the envelope to loss mass
to inner regions.
This will decrease substantially the
ratio $M_{\rm env}/M_{\rm env0}$ in equation (13), hence substantially
reducing the angular velocity as a result of the mass loss.
The single-star models for the bipolar Great Eruption, therefore,
do not directly address the question of what is
the main factor behind the bipolar structure.
{{{ There is, however, more to the single star model.
Smith et al.\ (2003a), for example, attribute the 5.5 year period
to angular momentum redistribution in the envelope.
The star eruptively expels large amount of mass and angular
momentum, such that its surface slows down.
The star then relaxes, and angular momentum diffuses outward.
After a period of 5.5 years, the surface rotates fast enough
for a new eruption to occur.
Although I see some problems with this scenario, it does,
however, show that the angular momentum evolution in a single
star scenario is more complicated than what I described
in previous sections. }}}

The goal of the present paper was to show that the origin of the fast
envelope rotation required by these models cannot be ignored.
{{{ Even when fast rotation occurs, it is not clear
how much it can affect instabilities and the mass loss process
(Stothers 1999) }}}
I used general considerations, rather than a specific model.
 (Most papers dealing with a fast rotating progenitor of $\eta$ Car
do not actually provide such a stellar model!)
The first problem is the initial fast rotation required. This in principle
can be provided by a relatively low mass companion which entered the
envelope of $\eta$ Car in the past. The more difficult problem to
overcome, as I showed in section 3 above, is to maintain a fast
envelope rotation during the Great Eruption itself.
On top of the angular momentum problem, the models are further
constrained by the huge momentum and kinetic energy of the mass
blown during the great eruption (section 2 above).
These, for example, rule out a model where most of the mass is blown
from the polar caps of the progenitor.

Binary companions, on the other hand, can relatively easily account
for the shaping of the Homunculus (Soker 2001), and the energy and
momentum of the Great Eruption (section 2 bove).
The basic process is that of a companion accreting from the primary
wind, forming an accretion disk, and blowing two opposite jets.
The energy source is the gravitational energy of the mass accreted
on the companion.
Comparing LBV's nebulae with other objects, e.g., PNs, strongly
suggests that binary interaction is indeed behind the shaping
of these non-spherical nebulae (e.g., O'Hara et al.\ 2003).
The binary model has another advantage: it incorporates $\eta$ Car
into a unified model explaining all objects having two low density
lobes with an equatorial waist between them; sometimes these are
termed bubble pair.
These objects include clusters of galaxies, where the bubbles are
X-ray deficient bubbles, symbiotic nebulae, and planetary nebulae
(see Soker 2003 a,b; 2004a). The shaping in all these objects is
via supersonic jets.
The jets are blown by accretion disks at a speed about equal to
the escape velocity from the accreting object (Livio 2000).

Following the analytical exploratory papers of the shaping problem
from a binary-model perspective (Soker 2001; 2003c,
and the present paper), the next step in understanding the shaping
of the Homunculus is to conduct a 3D numerical simulation of
the mass transfer from $\eta$ Car to its companion, including two
oppositely ejected jets, with possibly wide opening angle
(Soker 2004a), and including the orbital motion.

\bigskip

{\bf ACKNOWLEDGEMENTS}

{{{ I thank an anonymous referee for useful comments.  }}}
This research was supported in part by a grant from the Israel
Science Foundation.

\end{document}